\documentclass[fleqn,twoside]{article}
\usepackage{espcrc2}
\usepackage{latexsym}
\usepackage{euscript}

\usepackage{graphicx}
\usepackage[figuresright]{rotating}

\def\simge{\mathrel{%
   \rlap{\raise 0.511ex \hbox{$>$}}{\lower 0.511ex \hbox{$\sim$}}}}   
\def\simle{\mathrel{   
   \rlap{\raise 0.511ex \hbox{$<$}}{\lower 0.511ex \hbox{$\sim$}}}}   
\def\slashchar#1{\setbox0=hbox{$#1$}           
   \dimen0=\wd0                                 
   \setbox1=\hbox{/} \dimen1=\wd1               
   \ifdim\dimen0>\dimen1                        
      \rlap{\hbox to \dimen0{\hfil/\hfil}}      
      #1                                        
   \else                                        
      \rlap{\hbox to \dimen1{\hfil$#1$\hfil}}   
      /                                         
   \fi}                                         %

\def\simge{\mathrel{%
   \rlap{\raise 0.511ex \hbox{$>$}}{\lower 0.511ex \hbox{$\sim$}}}}   
\def\simle{\mathrel{   
   \rlap{\raise 0.511ex \hbox{$<$}}{\lower 0.511ex \hbox{$\sim$}}}}   
\def\slashchar#1{\setbox0=\hbox{$#1$}           
   \dimen0=\wd0                                 
   \setbox1=\hbox{/} \dimen1=\wd1               
   \ifdim\dimen0>\dimen1                        
      \rlap{\hbox to \dimen0{\hfil/\hfil}}      
      #1                                        
   \else                                        
      \rlap{\hbox to \dimen1{\hfil$#1$\hfil}}   
      /                                         
   \fi}

\newcommand{\AmS}{{\protect\the\textfont2
  A\kern-.1667em\lower.5ex\hbox{M}\kern-.125emS}}

\newcommand{\ba}{\begin{equation} \left\{ \begin{array}{lr}}
\newcommand{\ea}{\end{array} \right. \end{equation}}

\newcommand{\bea}{\begin{eqnarray}}
\newcommand{\eea}{\end{eqnarray}}
\newcommand{\beqa}{\begin{eqnarray}}
\newcommand{\eeqa}{\end{eqnarray}}
\newcommand{\be}{\begin{equation}}
\newcommand{\ee}{\end{equation}}
\newcommand{\beq}{\begin{equation}}
\newcommand{\eeq}{\end{equation}}

\newcommand{\MSbar}{\hbox{$\overline{MS}$\ }}

\title{Heavy quark masses from finite volume effects\thanks{Talk given at Lattice 2003}}

\author{Nazario Tantalo\address[ROME2]{Dipartimento di Fisica, Universit\`a di Roma ``Tor Vergata'', 
        V. R. Scientifica 1, I-00133 Rome, Italy}\address[INFN]{INFN Roma 2, V. R. Scientifica 1, I-00133 Rome, Italy}}
       
\begin{document}

\begin{abstract}
I discuss the results of a new calculation of the charm
and bottom quark masses in the quenched
approximation and in the continuum limit of lattice QCD. 
The work has been done by
the APE group at the ``Tor Vergata'' University \cite{deDivitiis:2003iy} making use of  
the step scaling method, previously introduced to deal with two scale
problems, that allows to take the continuum limit of the lattice data. 
We have computed the RGI quark masses and then we have connected the results to the
\MSbar scheme. 
The continuum numbers are $m_b^{RGI} = 6.73(16)$ GeV for the $b$--quark 
and $m_c^{RGI} = 1.681(36)$ GeV for 
the $c$--quark, corresponding respectively to $m_b^{\overline{MS}}(m_b^{\overline{MS}}) = 4.33(10)$ GeV
and $m_c^{\overline{MS}}(m_c^{\overline{MS}}) = 1.319(28)$ GeV. 
The latter result, in agreement with current estimates, is for us a check of the method.
\end{abstract}

\maketitle

\section{Introduction}

Quark masses are fundamental parameters of the QCD Lagrangian. Their accurate knowledge is required in order to 
give quantitative predictions of fundamental processes. A direct experimental measurement of quark masses is
not possible because of confinement, and their determination can only be inferred from a theoretical 
understanding of the hadron phenomenology. The calculations can be numerically performed through different 
strategies, depending upon the quark flavor and the available computational facilities. Accurate non--perturbative 
measurements of the $u$, $d$, $s$ and $c$ quark masses have been obtained from a straightforward comparison of hadron 
spectroscopy and lattice QCD predictions. 

The situation is different for the $b$ quark, with present
computers capabilities. In principle, the $b$ quark mass could be
extracted, similarly to lighter flavors, by looking 
at the heavy--light and heavy--heavy meson spectrum. 
However, heavy--light mesons are characterized by the presence of two 
different  scales, i.e. $\Lambda_{QCD}$, that sets the wavelengths of
the light quark, and the heavy $b$--quark mass.  
Managing these two scales in a naive way 
would require a very large lattice ($O(100^4)$ points).

The task of determining the $b$--quark mass has been faced in literature by resorting to some
approximations of the full theory, 
e.g. HQET on the lattice \cite{Gimenez:2000cj}, 
lattice NRQCD \cite{Gray:2002vk}, or QCD sum rules \cite{Bauer:2002sh,Battaglia:2002tm}. 
A novel approach recently introduced is based on the non--perturbative renormalization of the
static theory and its matching to QCD \cite{Sommer:2002en,Heitger:2003xg}. 

An alternative approach to the bottom quark physics, based on finite size scaling, has been proposed in a previous 
paper \cite{Guagnelli:2002jd}, where it has been applied to the
heavy--light meson decay constants
(see also \cite{deDivitiis:2003wy,pippo}). 
The main advantages of this
{\it step scaling method} (SSM) are that the entire computation is
performed with the relativistic QCD Lagrangian and that
the continuum limit can be taken, avoiding the unfeasible
direct calculation.
In order to implement the SSM, a finite size scheme is required, 
and we adopt the Schr\"odinger Functional (SF) as the most useful
framework.

\section{Step scaling functions and HQET}

The SSM has been designed in order to deal with two scale problems in lattice QCD \cite{Guagnelli:2002jd}
and it has been discussed in detail in \cite{deDivitiis:2003wy,deDivitiis:2003iy}.
The main assumption of the 
method is that the finite size effects affecting the meson masses, $M_{\{P,V\}}$ have a mild dependence upon 
variations of the heavy quark mass and are 
controllable from a numerical point of view. 
Finite size effects
can be obtained from 
\beqa
\sigma_P \left(L, m_1, m_2 \right) = \frac{\left.M_P\left(m_1, m_2 \right)\right|_{2L}}
{\left.M_P\left(m_1, m_2 \right)\right|_{L}}
\nonumber \\ \nonumber \\
\sigma_V \left(L, m_1, m_2 \right) = \frac{\left.M_V\left(m_1, m_2 \right)\right|_{2L}}
{\left.M_V\left(m_1, m_2 \right)\right|_{L}}
\label{eq:masssigmas}
\eeqa
for the pseudoscalar and vector meson masses respectively. \\
To validate the hypothesis
of low sensitivity upon the high energy scale, we can make use
of the HQET predictions on the heavy--light meson masses.
In the infinite volume the pseudoscalar and vector
meson masses have the following expansion in terms of the heavy--quark
mass:
\beq
M_X(m_h, m_l) = m_h + \bar{\Lambda}(m_l) + \frac{\alpha_X(m_l)}{m_h} + \dots
\label{eq:hqetiv}
\eeq 
where $X \in \{P,V\}$.
Assuming the contribution of the $1/m^2_h$ corrections to be negligible,
at finite volume one has
\beq
M_X(m_h, m_l, L) = m_h + \bar{\Lambda}_X(m_l,L) + \frac{\alpha_X(m_l,L)}{m_h}
\label{eq:hqetsv}
\eeq 
where $\bar{\Lambda}_X(m_l,L)$ depends
upon the spin of the meson state because of the contamination
of the excited states to the finite volume correlations.
Using eqs.~(\ref{eq:masssigmas}) and (\ref{eq:hqetsv})
we obtain the HQET predictions for the step
scaling functions of the heavy-light
meson masses
\beq
\sigma_X(L,m_h,m_l) = 1 + \frac{\sigma^{(0)}_X(m_l,L)}{m_h} 
+ \frac{\sigma^{(1)}_X(m_l,L)}{m_h^2}
\label{eq:sigmahqetsv}
\eeq
This result requires some considerations.
First we want to stress that in the infinite heavy--quark mass
limit the step scaling functions have to be exactly equal to one, 
$\sigma_X(L, m_l, m_h \rightarrow \infty) = 1$\footnote{We thank A.~Kronfeld 
and R.~Sommer for having pointed out this property of $\sigma_X$.}. 
This represents a strong constraint for the fits of the
heavy--quark mass dependence of the step scaling functions. \\
The second observation concerns the number of terms to be considered in eq.~(\ref{eq:sigmahqetsv}).
At order $O(1/m_h)$ one has
\beq
\sigma^{(0)}_X(m_l,L) = \bar{\Lambda}_X(m_l,2L) - \bar{\Lambda}_X(m_l,L)
\eeq
corresponding to the static approximation in
eq.~(\ref{eq:hqetsv}). 
By increasing the physical volume $L$, the difference between
$\bar{\Lambda}_X(m_l,2L)$ and $\bar{\Lambda}_X(m_l,L)$
decreases because the two quantities have to be equal
in the infinite volume limit, making
the heavy--quark mass expansion of the finite volume effects
rapidly convergent. 
The same arguments apply to the coefficient $\sigma^{(1)}_X(m_l,L)$ that has to
be considered when in the expansion of the meson masses, eq.~(\ref{eq:hqetsv}),
the order $O(1/m_h)$ is taken into account.
In our calculation we have performed the fits of
the step scaling functions considering the $O(1/m_h^2)$
term, $\sigma^{(1)}_X$, beyond the so called static evolution ({\it SE})
that retains $\sigma^{(0)}_X$ only.
We will also report for comparison the fits for the {\it SE}
that are anyway compatible. 

\begin{figure}[h]
\begin{center}
\includegraphics[width=\columnwidth]{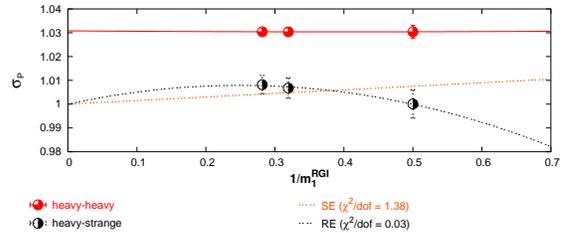}
\vskip -0.5cm
\caption{
The figure shows the continuum extrapolated step scaling functions $\sigma_P(L_0)$
as functions of $1/m_1^{RGI}$ with two possible fits.
The heavy extrapolations are shown only for the heavy--strange (Hs) set of data.}
\label{fig:A-S1-sigmas2}
\end{center}
\vskip -0.4cm
\end{figure}

\section{Results and Discussion}

\begin{figure}[h]
\begin{center}
\includegraphics[width=\columnwidth]{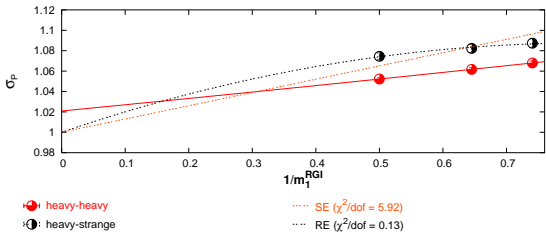}
\vskip -0.5cm
\caption{
The figure shows the continuum extrapolated step scaling functions $\sigma_P(2L_0)$
as functions of $1/m_1^{RGI}$  with two possible fits.
The heavy extrapolations are shown only for the heavy--strange (Hs) set of data.}
\label{fig:A-S2-sigmas2}
\end{center}
\vskip -0.4cm
\end{figure}

The physical numbers for the heavy meson spectrum have been 
obtained combining the results of a small
volume ($L_0 = 0.4$ fm) calculation, where the simulations have been performed
at the physical values of the heavy quark masses,
with the results of the step scaling functions according
to the identity:
\beq 
M_X(L_\infty) = M_X(L_0)\ \sigma_X(L_0)\ \sigma_X(2L_0)\ \dots
\label{eq:starting}
\eeq
The step scaling functions at the values of the heavy
quark masses simulated on the small volume
have been obtained by interpolation in the heavy--light case 
using eq.~(\ref{eq:sigmahqetsv}) while,
in the heavy--heavy case, the results are linearly extrapolated.
The plots of $\sigma_P(L_0)$ and $\sigma_P(2L_0)$ as functions of the inverse quark mass
are shown in figs.~\ref{fig:A-S1-sigmas2} and \ref{fig:A-S2-sigmas2}  
respectively.

Comparing the numerical heavy meson spectrum
with the experimental results \cite{PDBook} we have obtained
different determinations of the $b$--quark mass, depending upon
the physical state used as experimental input. \\
Within the quenched approximation, the determinations of the
quark masses coming from the heavy--heavy or from the heavy--light spectrum
in principle differ because the theory does not account for the fermion loops. 
The numbers we quote as final results are obtained by averaging the results 
for the heavy-heavy and the heavy-strange vector and pseudoscalar mesons 
and by keeping the typical error of a single case:
\beqa
m_b^{RGI} &=& 6.73(16) \mbox{ GeV} 
\nonumber \\
m_b^{\overline{MS}}(m_b^{\overline{MS}}) &=& 4.33(10) \mbox{ GeV}
\label{eq:mbfinal}
\eeqa
for the $b$--quark and
\beqa
m_c^{RGI} &=& 1.681(36) \mbox{ GeV} 
\nonumber \\
m_c^{\overline{MS}}(m_c^{\overline{MS}}) &=& 1.319(28) \mbox{ GeV}
\label{eq:mcfinal}
\eeqa
for the charm. 
The latter results compare favorably with the results of the direct computations \cite{Rolf:2002gu,PDBook}. 

Our error estimate includes both the \emph{statistical} error from the Monte Carlo
simulation as well as the \emph{systematic} error coming from the uncertainty on the lattice spacing corresponding at a given $\beta$ value
(we have used $r_0 = 0.5$ fm  \cite{Guagnelli:1998ud,Necco:2001xg,Guagnelli:2002ia})
and from the uncertainty on the renormalization constants.
The final errors on the continuum quantities, of the order of $2\%$ percent for the renormalization
constants and of about $1\%$ percent for the scale, are
added in quadrature and then linearly added to the statistical errors.
The evolution to the $\overline{MS}$ scheme has been done
using four--loop renormalization group equations.




\begin{thebibliography}{99}

\vskip 0.2cm

\bibitem{deDivitiis:2003iy}
G.~M.~de Divitiis, M.~Guagnelli, R.~Petronzio, N.~Tantalo and F.~Palombi,
arXiv:hep-lat/0305018.


\bibitem{Gimenez:2000cj}
V. Gimenez et~al.,
\newblock JHEP 03 (2000) 018, hep-lat/0002007.

\bibitem{Gray:2002vk}
HPQCD, A. Gray et~al.,
\newblock (2002), hep-lat/0209022.

\bibitem{Bauer:2002sh}
C.W. Bauer et~al.,
\newblock (2002), hep-ph/0210027.

\bibitem{Battaglia:2002tm}
M. Battaglia et~al.,
\newblock (2002), hep-ph/0210319.

\bibitem{Sommer:2002en}
R. Sommer,
\newblock (2002), hep-lat/0209162.

\bibitem{Heitger:2003xg}
J. Heitger, M. Kurth and R. Sommer,
\newblock (2003), hep-lat/0302019.

\bibitem{Guagnelli:2002jd}
M. Guagnelli et~al.,
\newblock Phys. Lett. B546 (2002) 237, hep-lat/0206023.

\bibitem{deDivitiis:2003wy}
G.~M.~de Divitiis, M.~Guagnelli, F.~Palombi, R.~Petronzio and N.~Tantalo,
arXiv:hep-lat/0307005.

\bibitem{pippo}
F. Palombi, these proceedings.

\bibitem{Guagnelli:1998ud}
ALPHA, M. Guagnelli, R. Sommer and H. Wittig,
\newblock Nucl. Phys. B535 (1998) 389, hep-lat/9806005.

\bibitem{Necco:2001xg}
S. Necco and R. Sommer,
\newblock Nucl. Phys. B622 (2002) 328, hep-lat/0108008.

\bibitem{Guagnelli:2002ia}
M. Guagnelli, R. Petronzio and N. Tantalo,
\newblock Phys. Lett. B548 (2002) 58, hep-lat/0209112.

\bibitem{PDBook}
Particle Data Group, K. Hagiwara et~al.,
\newblock Phys. Rev. D66 (2002) 010001.

\bibitem{Rolf:2002gu}
ALPHA, J. Rolf and S. Sint,
\newblock JHEP 12 (2002) 007, hep-ph/0209255.

\end{thebibliography}
\end{document}